\documentclass[conference]{IEEEtran}

\usepackage{hyphenat}
\usepackage{cite}
\ifCLASSINFOpdf
  \usepackage[pdftex]{graphicx}
  \graphicspath{{./figures/}{./graphs/}}
  \DeclareGraphicsExtensions{.pdf,.jpeg,.png}
\else
\fi
\ifCLASSOPTIONcompsoc
  \usepackage[caption=false,font=normalsize,labelfont=sf,textfont=sf]{subfig}
\else
  \usepackage[caption=false,font=footnotesize]{subfig}
\fi
\usepackage{url}
\usepackage{multirow}
\usepackage{amsmath}

\newcommand{\MySection}[1]{
  \vspace{-0.2ex}
  \section{#1}
  \vspace{-0.1ex}
}
\newcommand{\MySubsection}[1]{
  \vspace{-0.9ex}
  \subsection{#1}
  \vspace{-0.3ex}
}

\newcommand{\MyCaption}[1]{
  \vspace{-4.0ex}
  \caption{#1}
  \vspace{-1.0ex}
}
\usepackage{pifont}

\begin{document}

\title{
    %\vspace{-2.0ex}
    \fontsize{20}{20}\selectfont
    Optimized Broadcast for Deep Learning Workloads \\on Dense-GPU InfiniBand
Clusters: MPI or NCCL?
    %Design and Analysis of Large Message Broadcast for\\ Deep Learning Workloads on Dense Multi-GPU Clusters
    %Design and Analysis of Large Message Broadcast for\\ Deep Learning Workloads on Dense Multi-GPU Clusters
    %\vspace{-2.0ex}
    %\thanks{*This research is supported in part by U.S. Department of
    %        Energy grant \#DE-FC02-06ER25755;
    %        National Science Foundation grants \#CCF-0916302, 
    %        \#OCI-0926691 and \#CCF-0937842}
}

\author{Ammar Ahmad Awan, Ching-Hsiang Chu, Hari Subramoni, and Dhabaleswar K. (DK) Panda \\
%\vspace{1.6mm} 
\fontsize{10}{10}\selectfont\itshape
Department of Computer Science and Engineering, The Ohio State University\\
\vspace{-4.0ex}
\fontsize{9}{9}\selectfont\ttfamily\upshape
\{awan.10, chu.368\}@osu.edu, \{subramon, panda\}@cse.ohio-state.edu\\
}

\maketitle

% \fi
%\IEEEpeerreviewmaketitle

\thispagestyle{empty}

\pagenumbering{arabic}
\pagestyle{plain}

\begin{abstract}

Dense Multi-GPU systems have recently gained a lot of attention in the HPC
arena. Traditionally, MPI runtimes have been primarily designed for clusters with a
large number of nodes. However, with the advent of MPI+CUDA applications and
CUDA-Aware MPI runtimes like MVAPICH2 and OpenMPI, it has become important to
address efficient communication schemes for such dense Multi-GPU nodes. This
coupled with new application workloads brought forward by Deep Learning
frameworks like Caffe and Microsoft CNTK pose additional design constraints due
to very large message communication of GPU buffers during the training phase. In
this context, special-purpose libraries like NVIDIA NCCL have been proposed for
GPU-based collective communication on dense GPU systems. 
In this paper, we propose a pipelined chain (ring) design for the MPI\_Bcast 
collective operation along with an enhanced collective tuning framework in
MVAPICH2-GDR that enables efficient intra-/inter-node multi-GPU communication.
We present an in-depth performance landscape for the proposed MPI\_Bcast
schemes along with a comparative analysis of NVIDIA NCCL Broadcast and
NCCL-based MPI\_Bcast. The proposed designs for MVAPICH2-GDR enable
up to 14X and 16.6X improvement, compared to NCCL-based solutions, for intra- and 
inter-node broadcast latency, respectively. In addition, the proposed designs provide up to 7\%
improvement over NCCL-based solutions for data parallel training of the VGG network
on 128 GPUs using Microsoft CNTK.

\end{abstract}

\MySection{Introduction and Motivation}
\label{sec:intro}

Deep Learning (DL) is seeing a phenomenal growth spike where Computer Science
(CS) and Machine Learning (ML) communities alike are rapidly adopting Deep
Neural Networks (DNNs) to solve a huge array of classical problems.  In fact,
DNNs in specific, and DL, in general, has become one of the most disruptive
technologies of its time.  From Image Recognition to Speech Processing, various
ML areas are being completely taken over by DNN-based approaches. With Neural
Machine Translation (NMT)~\cite{nmt-model} based approaches, several years of
classical ML research has become obsolete in a matter of few
years~\cite{SennrichHB16}. Historically, machine learning experts used
to create and tune features that captured the ML task (e.g. image recognition)
by hand. However, DNNs have drastically changed this where the deep
convolutional network automatically extracts these features in a layered
fashion. At the core of DNN adoption are fast parallel hardware architectures
like NVIDIA GPUs and more recently, the Tensor Processing Units (TPUs) by
Google. In this context, the need for faster DNN training has pushed the current
hardware, system software, and communication middleware to their limits. 

DNN training is a compute intensive problem and many in the community have
stressed the need for better exploiting HPC resources to deal with
it~\cite{dean-nips-12, iandola2015firecaffe, s-caffe}. As DNN training becomes
too large to be performed on a single GPU (or a single compute node), several DL
toolkits and frameworks like Caffe~\cite{caffe}, TensorFlow~\cite{tensorflow},
and Microsoft Cognitive Toolkit (CNTK)~\cite{cntk} have resorted to
parallel/distributed training on multiple GPUs and compute nodes. To this end,
there are new opportunities and challenges for the High-Performance Computing
(HPC) community to rethink and enhance communication middleware like Message
Passing Interface (MPI) and enable low-latency and high-bandwidth communication.
The key idea in DNN training is that training parameters (or
weights) need to be exchanged among parallel trainers (or optimizers) before the
start of a training iteration. This can be achieved by using an MPI\_Bcast on
the training parameters. The details of DNN training are beyond the scope of
this paper and can be reviewed from~\cite{dl-review-nature}.

In this paper, we investigate efficient broadcast mechanisms used
today to realize data parallel training in current DL frameworks and 
address the following research challenges:

\begin{itemize}

\item What are the problems in adoption of MPI Collectives like Bcast for
large scale DNN training?

\item Why is the DL community developing in-house and special-purpose communication
runtimes like NCCL?

\item How can CUDA-Aware MPI runtimes be enhanced to better tackle large message
communication requirements brought forward by DL applications?

\item Can general-purpose MPI runtimes like MVAPICH2 be optimized to better deal 
with diverse communication requirements for DNNs like LeNet, AlexNet, ResNet, and VGG?

\item How can we design an efficient and scalable MPI\_Bcast implementation that
can meet or surpass the performance of a special-purpose library like NCCL?

\end{itemize}

\MySubsection{Contributions}
\label{sec:contrib}

To address these challenges, we first highlight the design complexity of
CUDA-Aware MPI runtimes like MVAPICH2-GDR~\cite{mvapich2} and how it enables end applications to
deliver the best performance across all message ranges for the candidate MPI\_Bcast
collective operation. To provide a holistic view, we compare the performance of
proposed enhanced designs for MPI\_Bcast, GPU-optimized NCCL Broadcast, and
MPI+NCCL integrated MPI\_Bcast~\cite{awan-eurompi16} using micro-benchmarks and
real-world DL frameworks. Along with the broad goal of developing an efficient
MPI\_Bcast design for DL workloads, we make the following key contributions in
this paper:

\begin{itemize}

\item Discuss the emerging trends in DL community and how the HPC community can
address these by advocating the adoption of MPI runtimes like MVAPICH2 for new
application areas like parallel DNN training

\item Propose and design new MPI\_Bcast algorithms that provide efficient
GPU-based communication across all message sizes (DL models) and multi-GPU nodes
of a dense GPU cluster

\item Provide performance insights for NCCL broadcast and its limitations for
certain DL model sizes and system architectures along with solutions at the
MPI runtime level that can overcome these limitations

\item Model and evaluate the performance of several algorithms and design
techniques to realize an efficient MPI\_Bcast that can provide better
performance than NCCL for both intranode and internode communication

\item Illustrate application level benefits for the proposed designs using
Microsoft CNTK and the popular VGG~\cite{vgg-paper} network.

\end{itemize}

With the proposed designs integrated into the MVAPICH2-GDR runtime, we report up 
to 14X and 16.6X lower latency (compared to NCCL-based solutions) for intra- and 
inter-node broadcast of GPU buffers. For large messages, our proposed designs
achieve comparable performance to NCCL-based solutions. At the application
level, proposed designs provide up to 7\% improvement over NCCL-based solutions
for data parallel training of the VGG network
on 128 GPUs using Microsoft CNTK. To the best of our knowledge, this is the first 
study that presents pure MPI-based designs that provide comparable or better
performance than NCCL-based solutions for Deep Learning workloads.

\MySection{The Design Space for GPU-based Collective Communication}
\label{sec:background}

We first discuss the emerging trend of dense multi-GPU systems like NVIDIA
DGX-1~\cite{dgx1}. Next, we describe the NCCL library and its importance for DL
workloads. Finally, we explain the concept of CUDA-Aware MPI along with the
limitations of NCCL integration in MVAPICH2-GDR.

\MySubsection{Dense Multi-GPU Systems}
Since GPU-enabled systems show promising capabilities to significantly accelerate
training process for deep learning~\cite{schmidhuber2015deep}, higher number of
GPUs is desired for the recent HPC systems. Thus, many dense multi-GPU systems
have been introduced. Cray CS-Storm cluster supercomputer~\cite{cray-cs-storm}
is equipped with InfiniBand interconnects and up to eight NVIDIA GPUs to enable
high-performance computation and communication for traditional HPC applications
as well as DL workloads. NVIDIA's new DGX-1V with eight NVIDIA Tesla V100 GPUs and 
300GB/s NVIDIA NVLINK2 interconnect provides 
96X faster training than Intel Xeon CPU-only server for ResNet50 training~\cite{dgx1}. 
The next generation
HPC systems, ~\textit{SUMMIT}~\cite{summit}, will be equipped with multiple NVIDIA Tesla
V100 GPUs, IBM Power9 processors, and InfiniBand EDR HCAs. 

\MySubsection{NVIDIA NCCL for GPU-based Collective Communication}
\label{sec:nccl}
NCCL (pronounced Nickel) is a recent GPU-based collective
communication library geared towards Deep Learning (DL) workloads. NVIDIA 
NCCL 1.3 is the latest open-source version available from the GitHub
site~\cite{nccl-github}. NCCL 2.0 announced at NVIDIA GTC will be released later this year. 
In this paper, we have used NCCL 1.3 for all the
experiments and design discussions. 

NCCL's API closely resembles the MPI interface and provides communication primitives for 
broadcast, all-gather, reduce, reduce-scatter, and all-reduce. Precisely, NCCL's goal is 
to provide fast communication of messages between GPUs in dense multi-GPU machines 
like the DGX-1. In this sense, NCCL is a
special-purpose GPU-optimized collective communication library. 
However, the current publicly released version is limited to only a single shared-memory
multi-GPU node where GPUs are attached to a PCIe level interconnect like a PLX
switch. This offers the ability to optimize data exchange at very fine
warp-level granularity as well as the exploitation of Peer-Access and CUDA kernels for
copying the data instead of cudaMemcpy. Despite being similar to MPI, NCCL's design goals and the target
platforms are different. MPI is geared towards efficient communication across thousands of nodes 
in a cluster while NCCL is optimized for dense multi-GPU systems. 

\MySubsection{CUDA-Aware MPI}
In recent years, accelerators like NVIDIA GPUs are widely adapted by the modern
HPC systems~\cite{top500}. As a result, extensions have been proposed for MPI
runtimes to support efficient communication between GPUs. 
Initially, without the capability of direct access of GPU memory, MPI applications
required explicit copying of GPU data to a staging
buffer on the host memory in order to push the data to the network. Similarly, a
data movement from CPU to GPU was needed after receiving the data on the host
through an MPI\_Recv operation. This significantly impacts performance as well
as productivity. Thus, several MPI libraries including
OpenMPI~\cite{openmpi} and MVAPICH2-GDR~\cite{mvapich2} provide
\textit{CUDA-Aware} MPI primitives to transparently perform such copy
operations. These CUDA-Aware MPI libraries significantly improve performance and
productivity for the CUDA-enabled applications.

The internals of a CUDA-Aware MPI runtime are designed to have many optimized GPU-based
point-to-point communication schemes
such as staging, pipelining, CUDA IPC, and GPUDirect RDMA (GDR) to provide the
best performance across various scenarios like intra-node, intra-socket, inter-node, and
several other communication paths. This enables efficient designs for
collectives like the MPI\_Bcast for direct exchange of GPU-resident data.

\MySubsection{Limitations of NCCL-integrated MPI Designs}

CUDA-Aware MPI runtimes like MVAPICH2-GDR are flexible enough to integrate
third-party libraries like NCCL. In this context, we designed and evaluated NCCL-based MPI\_Bcast 
designs in our earlier work~\cite{awan-eurompi16}. The hierarchical nature of
collective communication in MVAPICH2 allowed us to exploit NCCL for intranode
communication along with efficient and tuned designs for internode
communication. However, there are certain limitations of integrating NCCL and
its likes in MPI runtimes. These include dealing with CUDA-specific features
like stream creation and management, and NCCL communicator creation and management
in addition to the MPI communicators. In addition, for systems that do not have
peer-access for GPUs, multiple NCCL communicators might need to be created to
extract the best performance. In general, NCCL integration with MPI runtimes
might lead to very complicated designs. Thus, the proposed work is a step
towards achieving similar or better performance without utilizing NCCL.

\MySection{Performance Modeling of Existing Broadcast Algorithms}
\label{sec:model}

Broadcast algorithms have been well-studied in the literature over the
decades~\cite{Thakur:2005:OCC,Zhou:2015:ICPPW,Chiba:2007:CCGrid}.
However, the emergence of accelerators such as GPUs has significantly changed
this field of research.
There is a need to re-evaluate existing algorithms as well as explore newer
design techniques that take advantage of GPU-specific hardware features and
mechanisms for realizing an efficient implementation of these algorithms. In 
this section, we model and analyze a few well-known broadcast algorithms to
study their performance characteristics. Table~\ref{tab:key} provides the
notations used for the performance model throughout this paper.

\begin{table}[htbp]
    \MyCaption{Notations for the Analytical Model}
    \label{tab:key}
    \centering
\resizebox{0.8\columnwidth}{!}{
        \begin{tabular}{ | c | p{6.7cm} |}
        \hline
        Name & Description \\
        \hline \hline
        $M$ & Size of a message \\ \hline
        $C$ & Size of a chunk \\ \hline
        $B$ & Bandwidth of the link \\ \hline
        $B_{PCIe}$ & The PCIe link bandwidth available for transfers between CPU
and GPU \\ \hline
        $n$ & The number of nodes (or GPUs) \\ \hline
        $t_{s}$ & The startup time for initiating a single transfer \\ \hline
        \end{tabular}
    }
        \vspace{-1em}
\end{table}

\MySubsection{Fundamental Broadcast Algorithms}
\label{sec:fund-bcast}

We first analyze the most common broadcast algorithms in the
literature that can be applied for the GPU-based broadcast. 
Note that the specialized broadcast
schemes~\cite{Mamidala:IPDPS:2006,hoefler-cac07,Venkatech:2014:HiPC:gpu-mcast,Chu:2017:ICPP}
that require hardware assisted features like the InfiniBand hardware multicast
are beyond the scope of this paper. 

\noindent \textbf{Direct Algorithm}: A direct algorithm to broadcast the data
from the root to all other processes is to simply use a serialized loop of point-to-point communication calls. In MPI, such an algorithm will essentially be a
loop of MPI\_Send and MPI\_Recv calls. The approach can be modeled as:

\begin{equation}
    \label{eq:direct-gdr}
    T_{(Bcast\_Direct)}=n\times(t_s + \frac{M}{B})
\end{equation}

However, this is not used in practice because of its poor scalability w.r.t to
the number of nodes `n'.

\noindent \textbf{Chain and Ring Algorithms}: 

The Chain algorithm is essentially an optimization of the direct
algorithm. Instead of having a single sender, i.e., the root process, each
successive recipient of the data becomes a sender for the next process in the
chain. For rooted collectives like MPI\_Bcast, the chain is a logical ring of
communicating processes without a wrap-around between the last and first
process in the chain. For non-rooted collectives, the wrap-around for the last
and first process essentially makes it a ring. 

Eq.~\ref{eq:chain} models the cost for the Chain algorithm.

\begin{equation}
    \label{eq:chain}
    T_{(Bcast\_Chain)}=(n-1)\times(t_s + \frac{M}{B})
\end{equation}

\noindent \textbf{Knomial/Binomial Tree Algorithm}: As the scale of the HPC
clusters becomes larger and larger, scalability with respect to the number of nodes becomes 
a critical performance issue. Thus, the tree-based approaches have been proposed 
to further improve the scalability of broadcast. In the Knomial tree-based approach, 
the set of communicating processes is considered as a logical tree where the
root of the tree is the root of the Broadcast operation. In each communication step, 
root forwards data to one of its child nodes unless the data has already
been sent to it.  In each step, the non-leaf nodes, which have received data from its
parent, follow the same procedure to forward data to their children nodes. 
In the Knomial tree, the root has at most $\lceil\log _k n\rceil$ children
to maximize the overlap of communications. When \textit{K} equals to two, it becomes the
well-known binomial tree algorithm. The communication cost of Knomial tree-based
approach can be represented as follows.
\begin{equation}
    \label{eq:knomial}
    T_{(Bcast\_Knomial)}=\lceil\log _k n\rceil\times(t_s + \frac{M}{B})
\end{equation}
As can be seen in Eq.~\ref{eq:knomial}, the communication can be reduced to the
scale of $O(\log _kn)$, which is significantly lower than the approaches
mentioned previously, which are $O(n)$.
As the tree-based algorithms improve
the scalability of the broadcast operation, these are widely used in MPI runtimes
and many other collective operations are built on top of it.

\noindent \textbf{Scatter-Allgather Algorithm}: The Scatter-Allgather algorithm
is slightly non-intuitive yet very powerful when scalability with respect to the message
size `M' is of higher importance than scalability with respect to the number of nodes `n'.
The scatter-allgather schemes~\cite{scatter-allgather-1,scatter-allgather-2} to
realize a broadcast have been proposed to improve the performance for large-message 
broadcast operations. The idea is to use a Scatter operation, followed by an 
Allgather operation to achieve a bandwidth-optimal broadcast. Typically, a
binomial-tree based Scatter followed by a ring-based Allgather operation is used
to realize the broadcast operation. The cost of this algorithm can be
represented by Eq.~\ref{eq:scatter-ring-allgather}, which is described in~\cite{Thakur:2005:OCC}.

\begin{align}
    \label{eq:scatter-ring-allgather}
    T&_{(Bcast\_Scatter\_Ring\_Allgather)} \notag \\
    &= \lceil\log _2n\rceil\times t_s + \frac{(n-1)\times M}{n}\times\frac{1}{B} \notag \\
    &+ (n-1)\times(t_s + \frac{M}{n}\times\frac{1}{B}) \notag \\
                &= (\lceil\log _2n\rceil+n-1)\times t_s +
                2\times\frac{(n-1)\times M}{n}\times\frac{1}{B} 
\end{align}

However, efficient implementation of this algorithm may not be widely available
as conventional HPC applications lie at the lower end of the spectrum with
respect to the message size `M'.

\MySubsection{Discussion}
\label{sec:discussion}

Depending on the number of nodes and message size, these algorithms and
corresponding implementations in MPI runtimes are judiciously applied to 
obtain optimal performance for a specific message range and node count. Despite
the rich space of these algorithms, not all MPI implementations provide
efficient designs for these algorithms especially when it comes to CUDA-Aware
MPI runtimes. 

\MySection{Proposed Design Schemes for CUDA-Aware MPI\_Bcast}
\label{sec:design}

We first discuss some advanced schemes and techniques that build on top of
existing algorithms discussed in Section~\ref{sec:model}. Next, we present the
design of the proposed chain algorithm and how we enhance the optimization framework
in MVAPICH2-GDR to exploit this new design. Finally,
we provide an overview of the various designs and algorithms in MVAPICH2-GDR 
that enable us to achieve the best performance for the entire message range. 

\MySubsection{Advanced Pipelining Schemes for Optimized Broadcast}
\label{sec:design-pipeline}

The cost functions presented in Section~\ref{sec:fund-bcast} always 
considered the transfer for the entire message in each communication
step, except for the \textbf{Scatter-Allgather} algorithm. As the interconnect
bandwidth has improved over the years, algorithms have been proposed to take
better advantage of it by chunking and/or pipelining these transfers. To better utilize 
the network resources, i.e., to saturate the available bandwidth, pipelining
schemes that divide the message into multiple smaller
\textit{\textbf{chunks}} need to be explored. Issuing multiple non-blocking point-to-point
communication calls, i.e., MPI\_Isend, MPI\_Irecv, to allow overlap of these communications is one strategy to implement a pipelined broadcast that achieves
better bandwidth utilization. Classically, the chain (or ring) algorithm has
been considered an inefficient algorithm by MPI implementers. However, with the
advent of Deep Learning applications, very large message transfers and a relatively smaller number of nodes (GPUs) are becoming a new use-case for MPI
runtimes. Thus, the conventional intuition around the broadcast algorithms
needs to be revisited.

\MySubsection{Proposed Pipelined Chain Design}
\label{sec:design-proposed-chain}

We propose a CUDA-Aware pipelined chain design for MPI\_Bcast in MVAPICH2-GDR. 
To realize the pipelined chain design, the root process chunks the data and
starts pushing the chunks to its right neighbor in the logical chain of
processes. All non-root processes except the last process in the chain, receive
several chunks from their left neighbor and forward it to their right neighbor. The last
process in the chain receives the chunks from its left neighbor only and doesn't
need to forward them any further. 

%required $\frac{M}{C}$ steps to forward all the data chunks to its neighbor.
%Finally, $(n-2)$ more steps are required to allow last data chunk be traveled
%over
%all processes. One can also allow root to exit once the whole message is pushed
%to the network, thus, the following $(n-2)$ steps can be ignored in the
%perspecitve of root.
%

The cost model for the pipelined chain can be extended based on the earlier
discussed chain algorithm from Section~\ref{sec:model}.

\begin{align}
    \label{eq:chain-chunk}
        T_{(Bcast\_Chain\_Pipeline)}=(\frac{M}{C}&+(n-2))\times(t_s +
\frac{C}{B})
\end{align}

Pipelining schemes theoretically yield lower communication costs; however, it 
is always non-trivial to select the proper chunk size that yields optimal
performance for different message sizes and various architectures. For our
implementation, we experimentally determine the optimal chunk size and allow the
collective tuning infrastructure in the MVAPICH2-GDR runtime to select the correct 
chunk-size for best performance across a wide range of message sizes and
process counts. 

%Figure~\ref{fig:chain} illustrates the proposed pipelined chain 
%design. 

%\begin{figure}[htbp]
%\centering
%\includegraphics[width=0.8\columnwidth]{figures/bcast-algo.pdf}
%\caption{Proposed Pipelined Chain Design for MPI\_Bcast}
%\label{fig:chain}
%\end{figure}

%can also apply the pipelining scheme. In this
%case, the root requires $\frac{M}{C}$ steps to push the whole message to one of
%its
%child node and it has at most $\lceil\log _k n\rceil$ children nodes. Therefore,
%the communication cost can be formulated as Eq.~\ref{eq:knomial-pipeline}.
%%
%\begin{align}
%    \label{eq:knomial-pipeline}
%    T_{(Bcast\_Knomial\_Pipeline)}=(\frac{M}{C}\times\lceil\log _k
%n\rceil&)\times(t_s + \frac{C}{B})
%\end{align}

%\noindent \textbf{Implementation Techniques}
% STG-COLL: 
%  
%     - Shared memory collectives for intra-node 
%  
%     - Loopback Designs for Intra-node 
%     
%     - Host-staging for Inter-node 

\MySubsection{GPU-specific Optimizations for CUDA-Aware MPI\_Bcast}

Besides the traditional broadcast schemes mentioned in previous sections, many
optimized schemes have been proposed specifically for GPU-based broadcast operation. 

\noindent \textbf{Host-staging Scheme}: The performance issues with GDR reads for 
GPU-based point-to-point communications has been reported
in~\cite{potluri2013efficient}. Thus, the direct broadcast of GPU buffers can
suffer performance degradation for certain message ranges. A host-staging
scheme is used to avoid these bottlenecks. Essentially, root first moves the data
from GPU memory to the host memory. Then, a host-based broadcast is performed.
On the receiver side, data can be either directly written to the GPU memory by
using NVIDIA GDR feature or staged through the host.

A K-nomial tree design with host-staging can be modeled as follows.
\begin{align}
    \label{eq:knomial-gdr-stage}
    T_{(Bcast\_Knomial\_Staging)}=\frac{M}{B_{PCIe}}&+\lceil\log _k
n\rceil\times(t_s + \frac{M}{B})
\end{align}

Clearly, we can see that this design can only perform better when the first
part of the equation, i.e., the time for copying the data to the host, doesn't
become the dominant part of the equation. For large and very-large messages, the
staging can become very expensive. Hence, for the pipelined chain design, we do
not perform host-staging and use direct mechanisms like CUDA IPC for intranode
transfers when GPUs have peer-access and CUDA GDR for internode transfers when
appropriate.

\MySection{Performance Analysis}
\label{sec:results}

We first provide the details of the evaluation platform we have used to perform
all the experiments. Next, we provide a comprehensive performance comparison of
several MPI\_Bcast schemes in MVAPICH2-GDR, NCCL-integrated MVAPICH2, and NCCL
using a micro-benchmark as well as a real DNN training framework called Microsoft Cognitive
Toolkit (CNTK).

\MySubsection{Evaluation Testbed}
We used a Cray CS-Storm based GPU cluster called KESCH~\cite{cscs-kesch} for our
experiments. The cluster is located at the Swiss National Supercomputing Center. 
KESCH is a dense multi-GPU 12-node cluster. Each node has eight NVIDIA K-80 GK210GL GPUs. 
Thus, a total 16 CUDA devices per node are available. With a total of 192 GPUs for the 12 nodes 
and just 24 conventional CPUs, KESCH is a dense cluster. Each node has two
InfiniBand FDR HCAs. Thus, KESCH is a multi-rail IB cluster that provides an opportunity for advanced designs in MVAPICH2-GDR to exploit both the HCAs. 

\MySubsection{Intranode Performance Comparison (Micro-benchmark)}
The goal of this section is to provide a comparison of performance for 
NCCL and MVAPICH2-GDR. We used a single node of the KESCH cluster and compare the
performance of NCCL and MVAPICH2-GDR for 2, 4, 8, and 16 GPUs. For brevity, we
compared the proposed tuned version of MVAPICH2-GDR (labeled as \textit{MV2-GDR-Opt}) 
and NCCL (labeled as \textit{NCCL}). We note that we exploit
the ncclBroadcast (NCCL) to realize the intranode broadcast in the NCCL-integrated
MPI\_Bcast design. Thus, we only compare NCCL and MVAPICH2 in this section and
omit the numbers for NCCL-integrated MPI\_Bcast. However, we show
NCCL-integrated MPI\_Bcast numbers for the inter-node comparison in the next section
(Section~\ref{sec:internode-perf}).

\begin{figure*}[htbp]
\centering
\subfloat[2 GPUs (1 node)]
{\includegraphics[width=0.45\textwidth]
{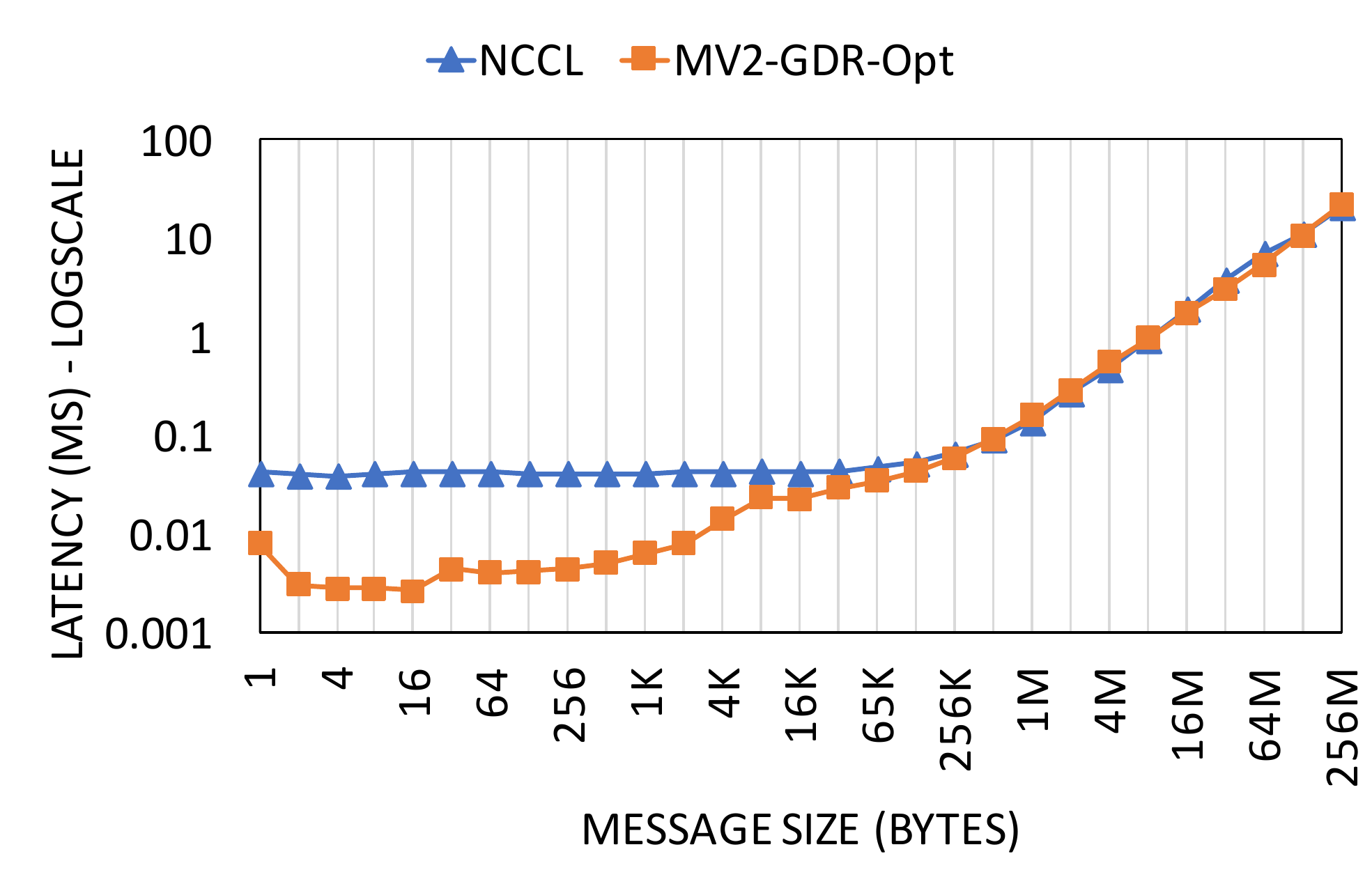}
\label{fig:163proc-small}}
\subfloat[4 GPUs (1 node)]
{\includegraphics[width=0.45\textwidth]
{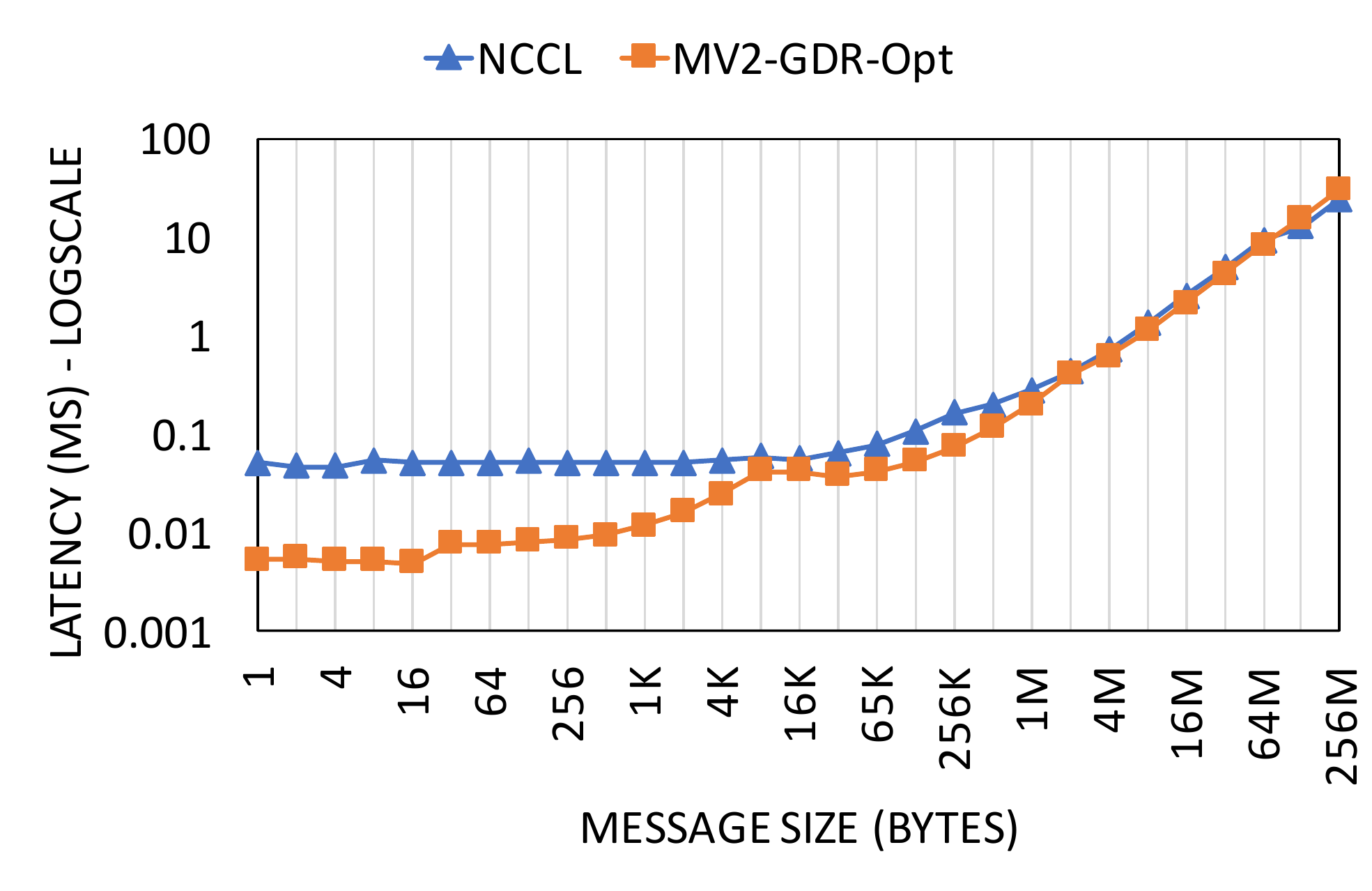}
\label{fig:163proc-med}}
\\
\subfloat[8 GPUs (1 node)]
{\includegraphics[width=0.45\textwidth]
{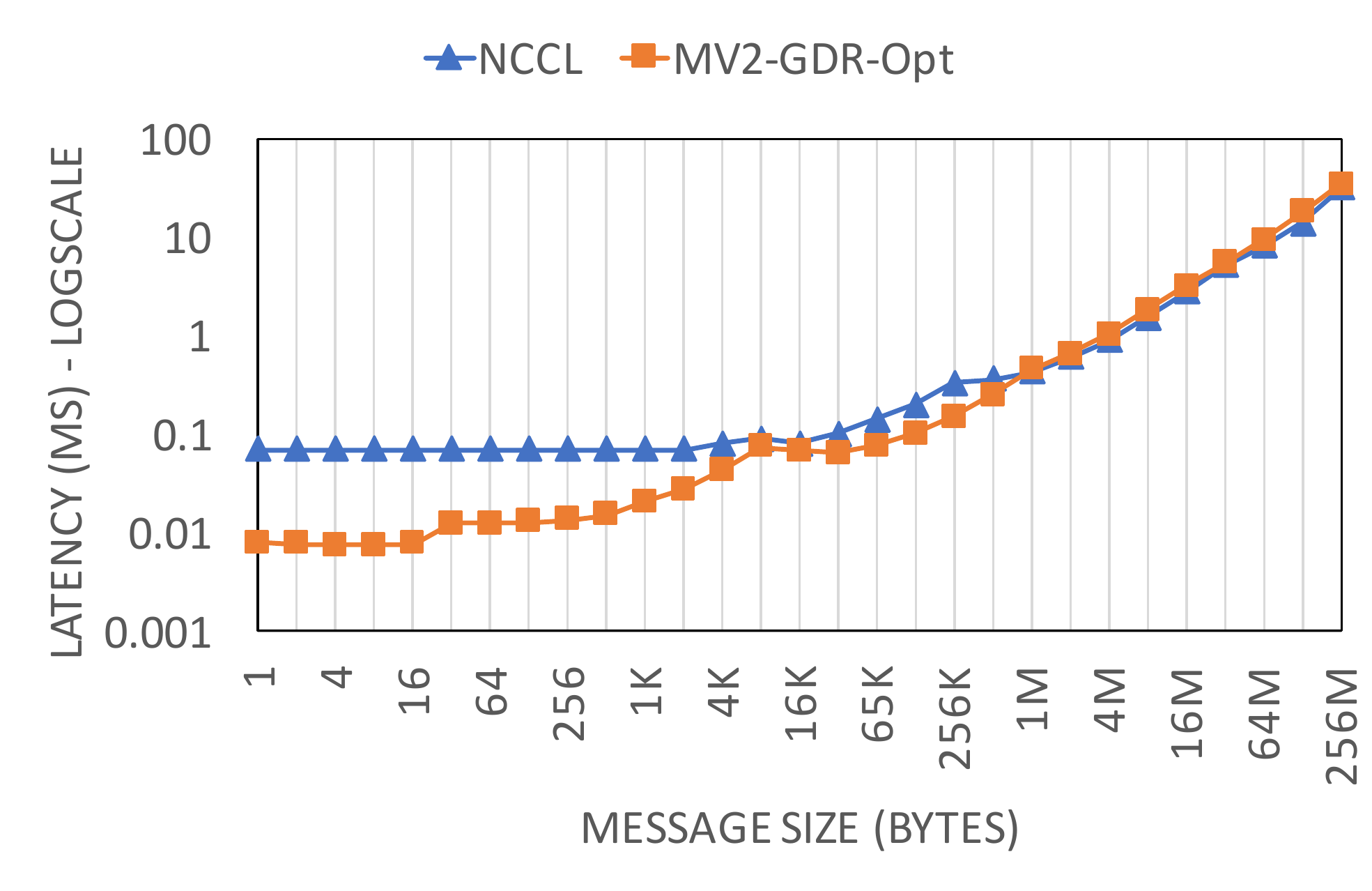}
\label{fig:163proc-med}}
\subfloat[16 GPUs (1 node)]
{\includegraphics[width=0.45\textwidth]
{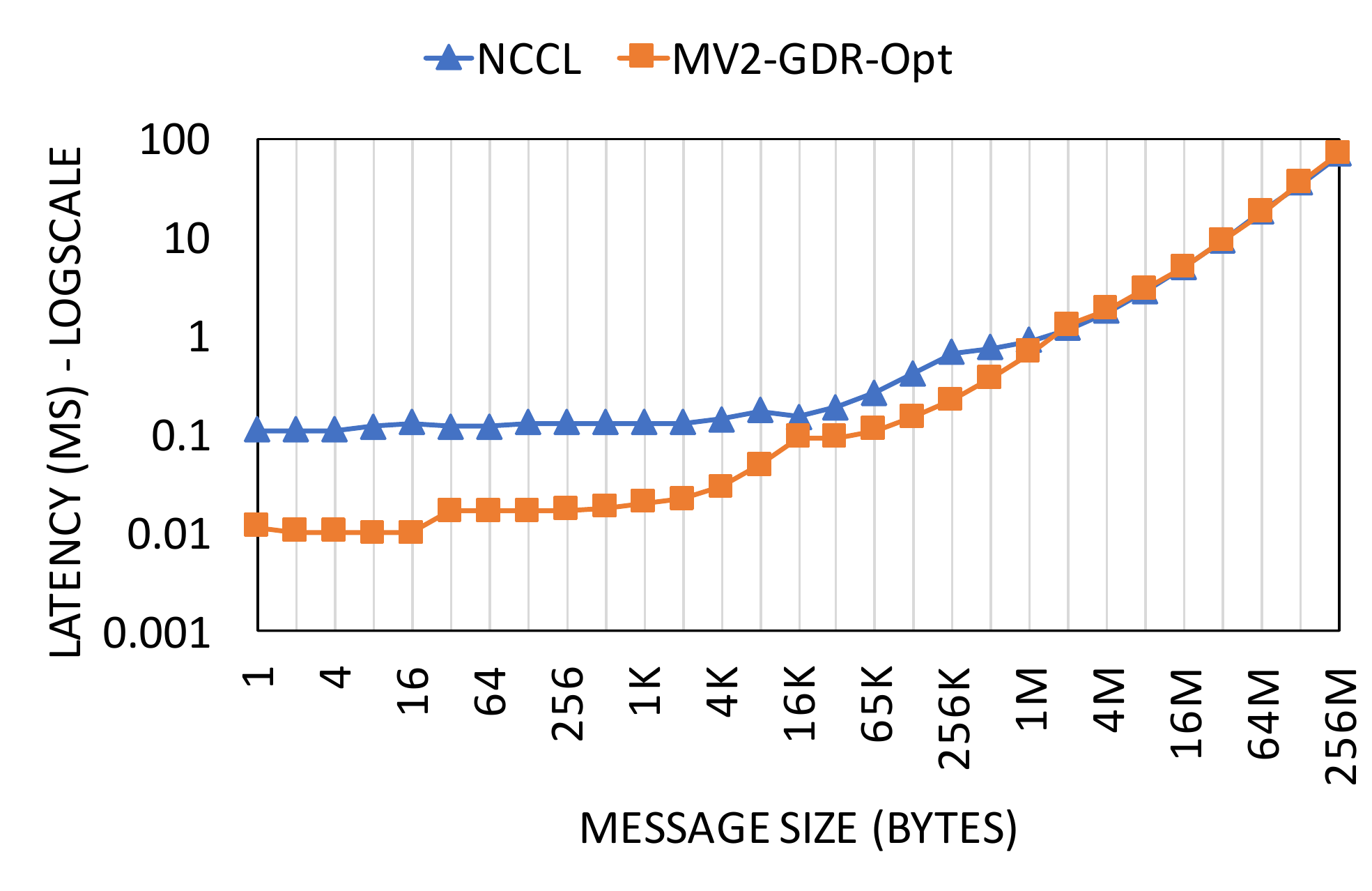}
\label{fig:163proc-med}}
\caption{Intranode Performance Comparison of NCCL and MVAPICH2\hyp{}GDR\hyp{}Optimized}
\label{fig:intra-results}
\end{figure*}

Figure~\ref{fig:intra-results} shows the intra-node (up to 16 GPUs) performance using the CUDA-Aware osu\_bcast
benchmark. For small to medium messages, i.e., up to 8K bytes, we observed up to
14X, 10.6X, 9.4X, and 13X lower latency in MPI\_Bcast of MVAPICH2-GDR compared to
NCCL Broadcast for 2, 4, 8, and 16 GPUs, respectively. This is because of the advanced
point-to-point designs available in MVAPICH2-GDR that allow efficient workarounds
for various bottlenecks like the GDR read bottleneck across
sockets~\cite{potluri2013efficient}. In addition, \textit{MV2-GDR-Opt} exploits pipelined CUDA IPC
designs for large messages to better utilize the available bandwidth. 
These optimized schemes cannot be done for special-purpose
libraries like NCCL and hence we see this degradation for small and medium
message ranges. For large and very large message range, we see that NCCL provides
scalable performance. At the same time, our proposed pipelined chain designs in MVAPICH2-GDR
allow us to achieve similar or better performance essentially alleviating the
need to resort to NCCL augmented broadcast designs proposed in~\cite{awan-eurompi16}.

\MySubsection{Internode Performance Comparison (Micro-benchmark)}
\label{sec:internode-perf}

As mentioned earlier, NCCL 1.x series only works for a single node so we cannot
directly compare the performance of NCCL for the internode case. However, our earlier
work on NCCL-based broadcast~\cite{awan-eurompi16} allows us to provide a
comprehensive internode performance comparison as well.

\begin{figure*}[htbp]
\centering
\subfloat[64 GPUs (4 nodes)]
{\includegraphics[width=0.45\textwidth]
{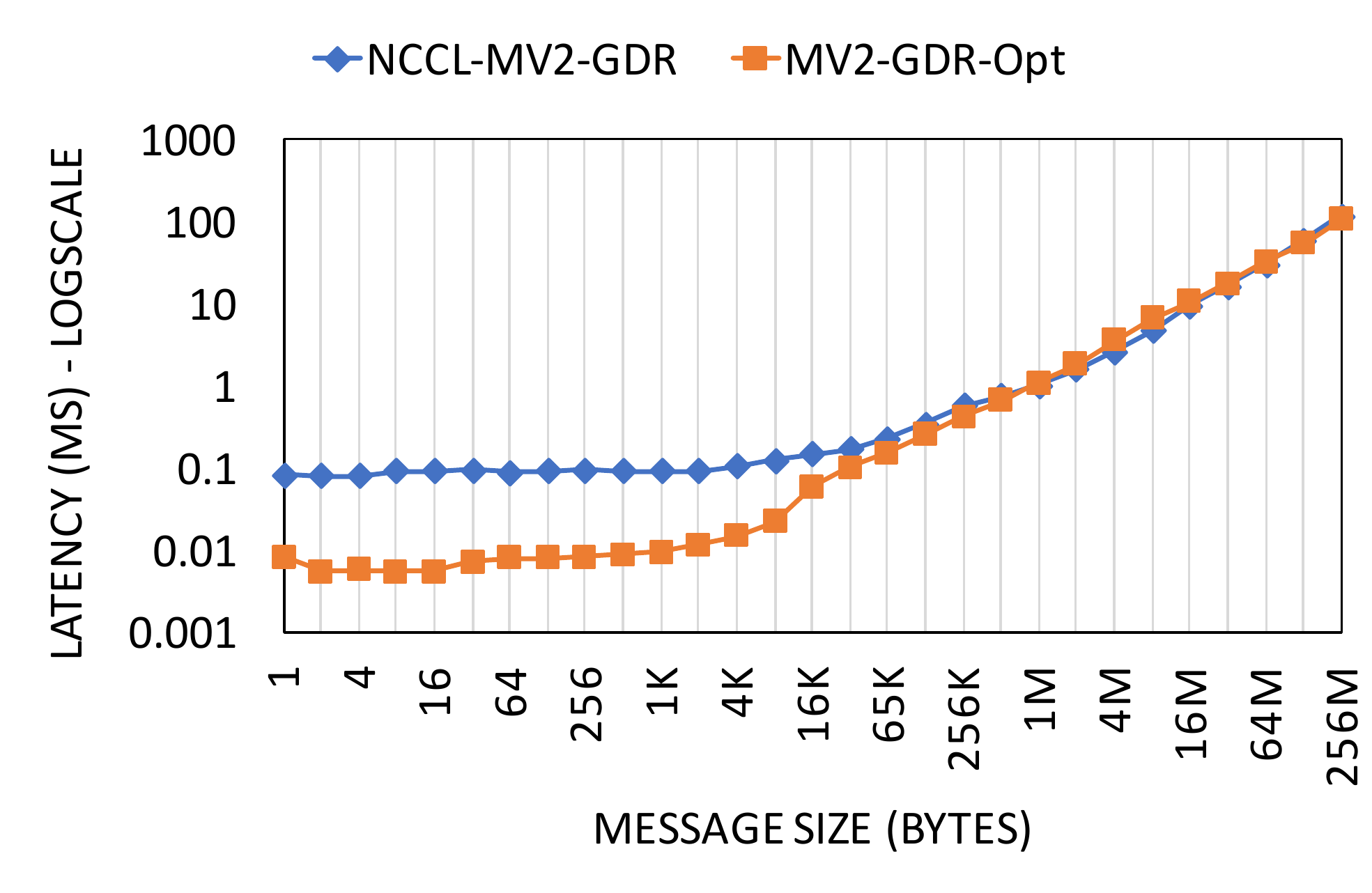}
\label{fig:163proc-small}}
\subfloat[128 GPUs (8 nodes)]
{\includegraphics[width=0.45\textwidth]
{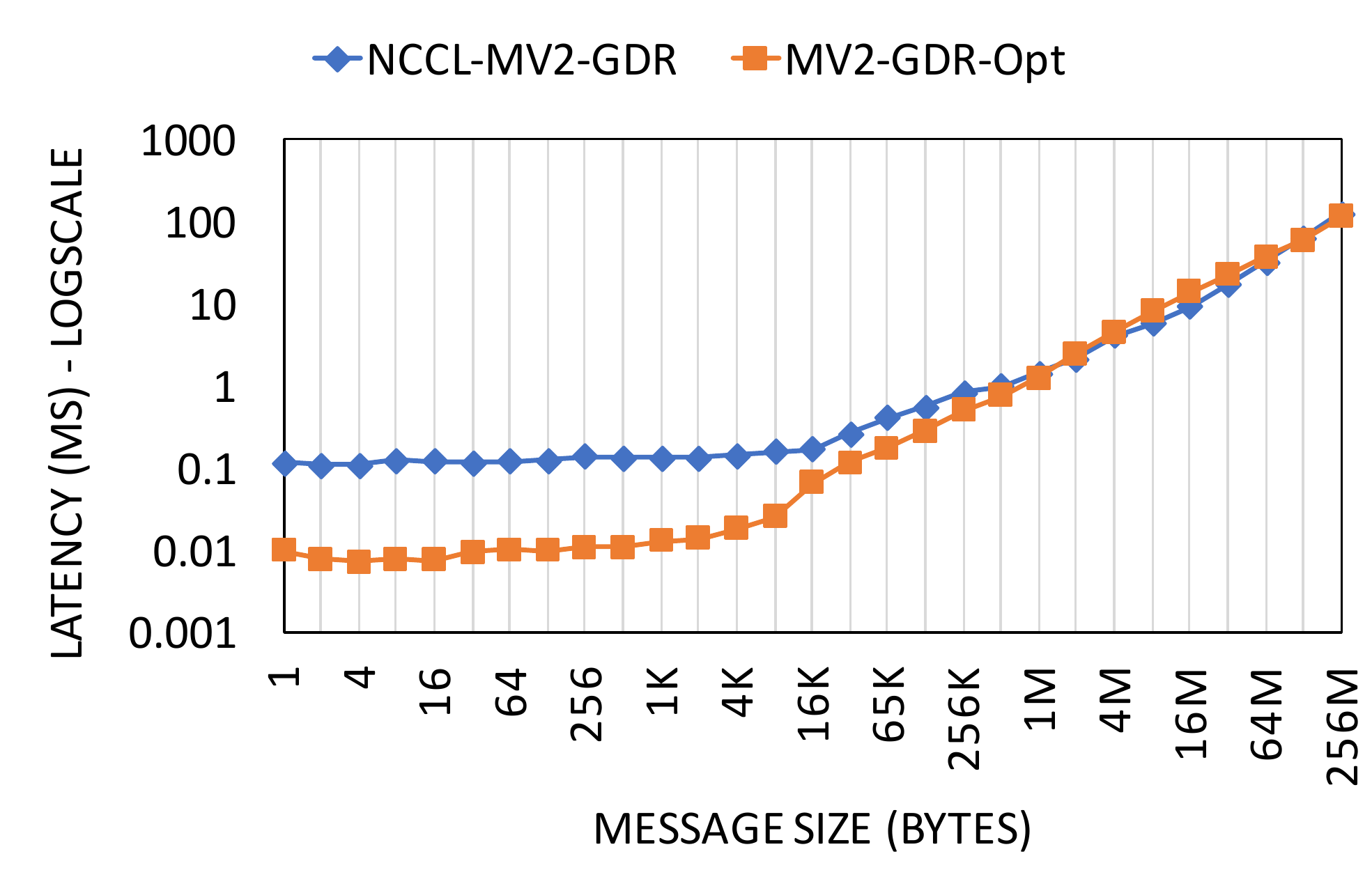}
\label{fig:163proc-med}}
\\
\caption{Internode Performance Comparison of NCCL-integrated MVAPICH2 and MVAPICH2\hyp{}GDR\hyp{}Optimized}
\label{fig:internode-result}
\end{figure*}

We compared two cases: the proposed tuned version of MVAPICH2-GDR (labeled
\textit{MV2-GDR-Opt}) and the NCCL-integrated MPI\_Bcast~\cite{awan-eurompi16}
(labeled \textit{NCCL-MV2-GDR}).
Several internode performance enhancements like the SGL-based designs~\cite{Shi:2014:HiPC} 
that take advantage of IB features like Scatter-Gather lists help us provide
excellent small message internode performance. As shown in
Figure~\ref{fig:internode-result}, MV2-GDR-Opt provides up to 16.4X and 16.6X
improvement, respectively, on 64 and 128 GPUs, for the small and medium message range. 
For large and very large messages, the
proposed designs are able to achieve comparable performance to the NCCL-based
designs. Thus, it is clear that MPI\_Bcast can provide the desired performance
for the various type of workloads including DL workloads. To better illustrate the
usefulness of these designs, we evaluate the performance of data parallel DNN
training using Microsoft Cognitive Toolkit (CNTK). The details follow in the
next subsection.

\MySubsection{Application Level Performance Comparison using CNTK}

\begin{figure}[htbp]
\centering
\includegraphics[width=0.8\columnwidth]
{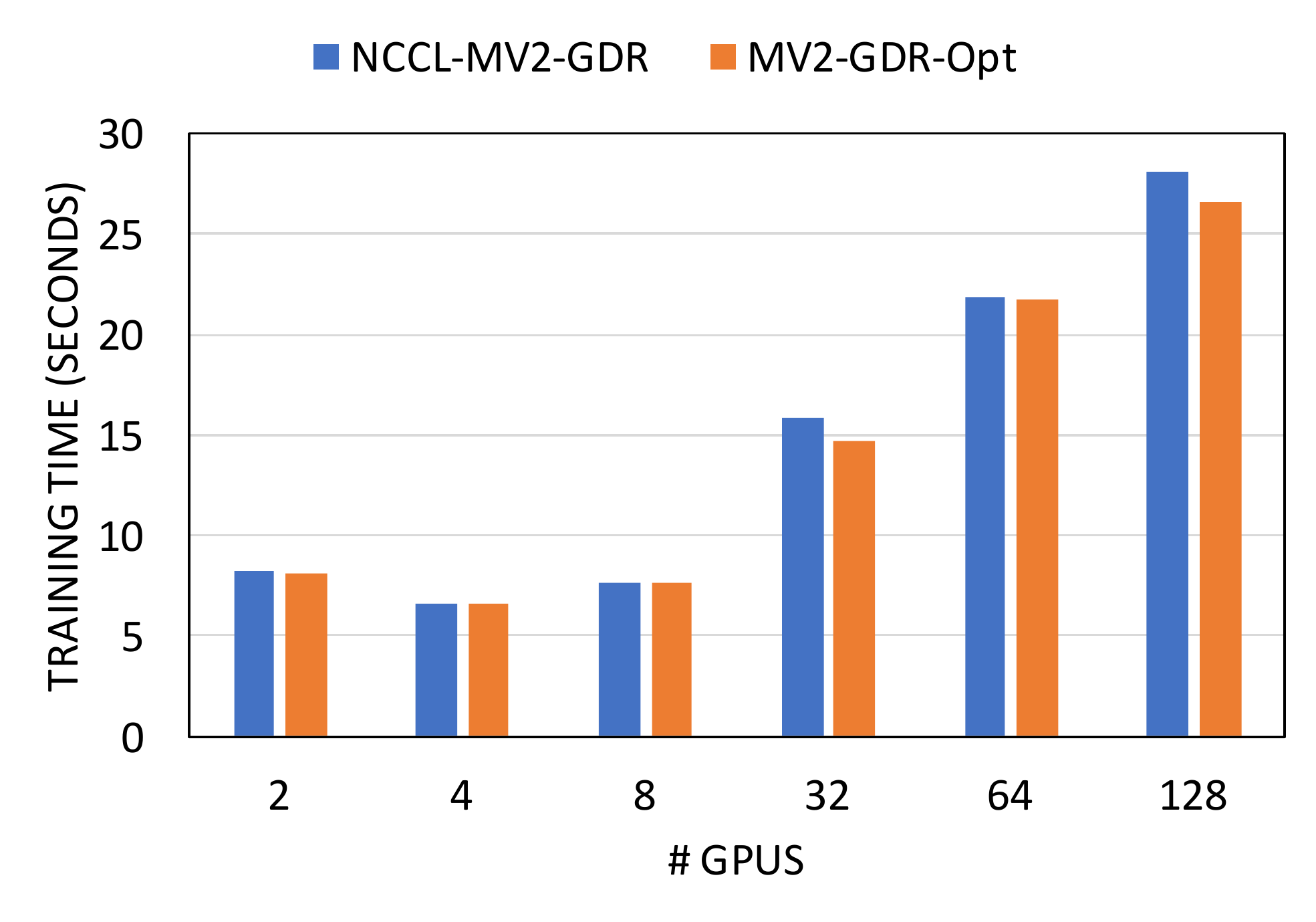}
\caption{Performance Comparison of NCCL-integrated MVAPICH2 and
MVAPICH2\hyp{}GDR\hyp{}Optimized for VGG Training with Microsoft CNTK}
\label{fig:cntk-results}
\end{figure}

We now provide a comprehensive view of the benefits that can be observed for a
real DNN training framework. There are several frameworks that can be utilized
but we used a CUDA-Aware adaptation of the CNTK framework called
CA-CNTK~\cite{dip:cloudcom16}. CA-CNTK uses CUDA-Aware MPI\_Bcast for the exchange 
of training parameters (or weights) throughout the training process. 
Figure~\ref{fig:cntk-results} shows the performance comparison of training time
using the VGG model for NCCL-integrated MVAPICH2 (labeled \textit{NCCL-MV2-GDR}) and
the proposed optimized version of MVAPICH2-GDR (labeled MV2-GDR-Opt). As can be seen
from Figure~\ref{fig:cntk-results}, MV2-GDR-Opt yields 7\% reduction of training
time on 32 GPUs, and matches or beats the performance
of NCCL-MV2-GDR for all other cases. This is because the broadcast operation used in
VGG training uses a mix of message sizes including some small and medium and
mostly large messages. 

This result illustrates that an optimized MPI runtime can offer better
performance than NCCL/NCCL-based approaches even when the
main requirement of an application (the VGG model in this case) is large-message 
communication. VGG model due to its large number of parameters forces large-message 
communication. Thus, the performance benefits observed for MV2-GDR-Opt on micro-benchmarks in the
small/medium message range do not exhibit significant benefits at the application level.
However, it is pertinent to note that CNTK divides the communication based on the process
count so the message-sizes can vary considerably and thus the proposed designs 
provide some performance benefit (7\%) for VGG. We expect the benefits to
increase for other models like GoogLeNet~\cite{googlenet} that have lesser
number of parameters and thus a small/medium message communication requirement. 

\MySection{Related Work}
\label{sec:related}

An efficient broadcast design was presented by Liu et al.
in~\cite{Liu:2004:IPDPS:ib-mcast} that takes advantage of
InfiniBand features like Hardware Multicast. The motivation of this work 
is similar to our proposed work, i.e. to optimize and redesign MPI\_Bcast by exploiting new design schemes for
newer architectures. Kandalla et. al presented designs to optimize MPI broadcast and reductions for
Intel Many Integrated Cores (MIC) in~\cite{kandalla-bcast}. 

Chu et al.~\cite{Chu:2017:ICPP} proposed an IB-Multicast based Broadcast for
streaming and DL applications. Zhou et. al proposed an optimized broadcast for large message sizes
in~\cite{Zhou:2015:ICPPW}. A slightly orthogonal effort to optimize intra-node
communication by exploiting
shared-memory programming and MPI one-sided features was presented
in~\cite{Hoefler:2012:EuroMPI}. Awan et al.~\cite{awan-eurompi16} described the
NCCL-integrated MPI\_Bcast designed for DL workloads. The main difference
between NCCL-integrated MPI\_Bcast and our proposed design is that it alleviates
the need to use NCCL or any other external library to accelerate the intranode
communication. Instead, we propose that efficient implementation of MPI\_Bcast
in the MVAPICH2 runtime can meet or exceed the performance offered by NCCL and
its likes.

\MySection{Conclusion and Future Work}
\label{sec:conclusion}

As we enter the era of next-generation AI systems powered by Deep Learning and
Deep Neural Networks, we see an increased exploration around communication
runtimes. Industry developments like NVIDIA NCCL is a good
example of such investigations. However, we propose that communication
runtimes like MVAPICH2 and other MPI implementations provide an excellent and
rigorous foundation that can be exploited to support communication with special
requirements. As such, there are no fundamental limitations that should sway the
researchers to not explore new designs in MPI runtimes to support DL workloads.
In this paper, we present this case using an in-depth investigation of
MPI\_Bcast as a candidate collective that is important for parallel DNN training. 
Using micro-benchmark and a real DNN training framework, we highlighted that
MPI\_Bcast designed and tuned efficiently can provide up to 16.6X speedup at the
micro-benchmark level and up to 7\% improvement for training the VGG network on
128 GPUs using Microsoft CNTK. We plan to make these solutions available in the next MVAPICH2-GDR release. We 
also plan to extend this support for other collectives like MPI\_Reduce and MPI\_Allreduce 
to support the full spectrum of parallel DNN training.

\section*{Acknowledgment}

This research is supported in part by National Science Foundation grants
\#CCF-1565414 and \#CNS-1513120.



%\bibliographystyle{./IEEEtranS}

%\bibliography{./bibs/header,./bibs/bibfile,./bibs/ching}
%\bibliography{IEEEabrv,./sample}

\begin{thebibliography}{10}
\providecommand{\url}[1]{#1}
\csname url@samestyle\endcsname
\providecommand{\newblock}{\relax}
\providecommand{\bibinfo}[2]{#2}
\providecommand{\BIBentrySTDinterwordspacing}{\spaceskip=0pt\relax}
\providecommand{\BIBentryALTinterwordstretchfactor}{4}
\providecommand{\BIBentryALTinterwordspacing}{\spaceskip=\fontdimen2\font plus
\BIBentryALTinterwordstretchfactor\fontdimen3\font minus
  \fontdimen4\font\relax}
\providecommand{\BIBforeignlanguage}[2]{{%
\expandafter\ifx\csname l@#1\endcsname\relax
\typeout{** WARNING: IEEEtranS.bst: No hyphenation pattern has been}%
\typeout{** loaded for the language `#1'. Using the pattern for}%
\typeout{** the default language instead.}%
\else
\language=\csname l@#1\endcsname
\fi
#2}}
\providecommand{\BIBdecl}{\relax}
\BIBdecl

\bibitem{cscs-kesch}
``{KESCH: Cray CS-Storm System},''
  \url{http://www.cscs.ch/computers/kesch_escha/index.html}.

\bibitem{cntk}
``{CNTK},'' \url{http://www.cntk.ai/}, 2015, [Online; accessed April-2016].

\bibitem{tensorflow}
M.~Abadi, A.~Agarwal, P.~Barham, E.~Brevdo, Z.~Chen, C.~Citro, G.~S. Corrado,
  A.~Davis, J.~Dean, M.~Devin \emph{et~al.}, ``{TensorFlow: Large-Scale Machine
  Learning on Heterogeneous Systems, 2015},'' \emph{Software available from
  tensorflow. org}.

\bibitem{awan-eurompi16}
A.~A. Awan, K.~Hamidouche, A.~Venkatesh, and D.~K. Panda, ``{Efficient Large
  Message Broadcast using NCCL and CUDA-Aware MPI for Deep Learning},'' in
  \emph{Proceedings of the 23rd European MPI Users' Group Meeting}.\hskip 1em
  plus 0.5em minus 0.4em\relax ACM, 2016, pp. 15--22.

\bibitem{s-caffe}
A.~A. Awan, K.~Hamidouche, J.~M. Hashmi, and D.~K. Panda, ``{S-Caffe:
  Co-designing MPI Runtimes and Caffe for Scalable Deep Learning on Modern GPU
  Clusters},'' in \emph{Proceedings of the 22Nd ACM SIGPLAN Symposium on
  Principles and Practice of Parallel Programming}, ser. PPoPP '17.\hskip 1em
  plus 0.5em minus 0.4em\relax ACM, 2017, pp. 193--205.

\bibitem{dip:cloudcom16}
D.~S. Banerjee, K.~Hamidouche, and D.~K. Panda, ``{Re-Designing CNTK Deep
  Learning Framework on Modern GPU Enabled Clusters},'' in \emph{2016 IEEE
  International Conference on Cloud Computing Technology and Science
  (CloudCom)}, Dec 2016, pp. 144--151.

\bibitem{scatter-allgather-1}
M.~Barnett, L.~Shuler, R.~van~de Geijn, S.~Gupta, D.~G. Payne, and J.~Watts,
  ``{Interprocessor collective communication library (InterCom)},'' in
  \emph{Proceedings of IEEE Scalable High Performance Computing Conference},
  May 1994, pp. 357--364.

\bibitem{Chiba:2007:CCGrid}
T.~Chiba, T.~Endo, and S.~Matsuoka, ``{High-Performance MPI Broadcast Algorithm
  for Grid Environments Utilizing Multi-lane NICs},'' in \emph{Seventh IEEE
  International Symposium on Cluster Computing and the Grid (CCGrid '07)}, May
  2007, pp. 487--494.

\bibitem{Chu:2017:ICPP}
C.-H. Chu, X.~Lu, A.~A. Awan, H.~Subramoni, J.~Hashmi, B.~Elton, and D.~K.
  Panda, ``{Efficient and Scalable Multi-Source Streaming Broadcast on GPU
  Clusters for Deep Learning},'' in \emph{46th International Conference on
  Parallel Processing (ICPP-2017)}, Aug 2017, [To appear].

\bibitem{cray-cs-storm}
\BIBentryALTinterwordspacing
{Cray}, ``{CS-STORM GPU-ACCELERATED CLUSTER SUPERCOMPUTER},'' {Accessed:
  \today}. [Online]. Available:
  \url{http://www.cray.com/products/computing/cs-series/cs-storm}
\BIBentrySTDinterwordspacing

\bibitem{dean-nips-12}
\BIBentryALTinterwordspacing
J.~Dean, G.~Corrado, R.~Monga, K.~Chen, M.~Devin, M.~Mao, M.~aurelio Ranzato,
  A.~Senior, P.~Tucker, K.~Yang, Q.~V. Le, and A.~Y. Ng, ``Large scale
  distributed deep networks,'' in \emph{Advances in Neural Information
  Processing Systems 25}, F.~Pereira, C.~J.~C. Burges, L.~Bottou, and K.~Q.
  Weinberger, Eds.\hskip 1em plus 0.5em minus 0.4em\relax Curran Associates,
  Inc., 2012, pp. 1223--1231. [Online]. Available:
  \url{http://papers.nips.cc/paper/4687-large-scale-distributed-deep-networks.pdf}
\BIBentrySTDinterwordspacing

\bibitem{hoefler-cac07}
T.~Hoefler, C.~Siebert, and W.~Rehm, ``{A Practically Constant-time MPI
  Broadcast Algorithm for Large-scale InfiniBand Clusters with Multicast},'' in
  \emph{Proceedings of the 21st IEEE International Parallel \& Distributed
  Processing Symposium (CAC'07 Workshop)}, Mar. 2007, p. 232.

\bibitem{Hoefler:2012:EuroMPI}
T.~Hoefler, J.~Dinan, D.~Buntinas, P.~Balaji, B.~W. Barrett, R.~Brightwell,
  W.~Gropp, V.~Kale, and R.~Thakur, ``{Leveraging MPI's One-sided Communication
  Interface for Shared-memory Programming},'' in \emph{Proceedings of the 19th
  European Conference on Recent Advances in the Message Passing Interface},
  ser. EuroMPI'12.\hskip 1em plus 0.5em minus 0.4em\relax Berlin, Heidelberg:
  Springer-Verlag, 2012, pp. 132--141.

\bibitem{iandola2015firecaffe}
F.~N. Iandola, K.~Ashraf, M.~W. Moskewicz, and K.~Keutzer, ``{FireCaffe:
  Near-Linear Acceleration of Deep Neural Network Training on Compute
  Clusters},'' \emph{arXiv preprint arXiv:1511.00175}, 2015.

\bibitem{caffe}
Y.~Jia, E.~Shelhamer, J.~Donahue, S.~Karayev, J.~Long, R.~Girshick,
  S.~Guadarrama, and T.~Darrell, ``{Caffe: Convolutional Architecture for Fast
  Feature Embedding},'' \emph{arXiv preprint arXiv:1408.5093}, 2014.

\bibitem{kandalla-bcast}
K.~Kandalla, A.~Venkatesh, K.~Hamidouche, S.~Potluri, D.~Bureddy, and D.~K.
  Panda, ``{Designing Optimized MPI Broadcast and Allreduce for Many Integrated
  Core (MIC) InfiniBand Clusters},'' in \emph{2013 IEEE 21st Annual Symposium
  on High-Performance Interconnects}, Aug 2013, pp. 63--70.

\bibitem{dl-review-nature}
\BIBentryALTinterwordspacing
Y.~LeCun, Y.~Bengio, and G.~Hinton, ``Deep learning,'' \emph{Nature}, vol. 521,
  no. 7553, pp. 436--444, 05 2015. [Online]. Available:
  \url{http://dx.doi.org/10.1038/nature14539}
\BIBentrySTDinterwordspacing

\bibitem{Liu:2004:IPDPS:ib-mcast}
J.~Liu, A.~R. Mamidala, and D.~K. Panda, ``{Fast and Scalable MPI-level
  Broadcast using InfiniBand's Hardware Multicast Support},'' in \emph{Parallel
  and Distributed Processing Symposium, 2004. Proceedings. 18th International},
  April 2004, p.~10.

\bibitem{Mamidala:IPDPS:2006}
A.~R. Mamidala, L.~Chai, H.-W. Jin, and D.~K. Panda, ``{Efficient SMP-aware
  MPI-level Broadcast over InfiniBand's Hardware Multicast},'' in
  \emph{Proceedings 20th IEEE International Parallel Distributed Processing
  Symposium}, April 2006, p.~8.

\bibitem{top500}
H.~Meuer, E.~Strohmaier, J.~Dongarra, and H.~Simon, ``{TOP 500 Supercomputer
  Sites},'' http://www.top500.org.

\bibitem{mvapich2}
{MVAPICH2: MPI over InfiniBand, 10GigE/iWARP and RoCE},
  https://mvapich.cse.ohio-state.edu/.

\bibitem{nmt-model}
\BIBentryALTinterwordspacing
G.~Neubig, ``{Neural Machine Translation and Sequence-to-sequence Models: {A}
  Tutorial},'' \emph{CoRR}, vol. abs/1703.01619, 2017. [Online]. Available:
  \url{http://arxiv.org/abs/1703.01619}
\BIBentrySTDinterwordspacing

\bibitem{dgx1}
\BIBentryALTinterwordspacing
{NVIDIA}, ``{DGX-1: Essential Instrument of AI Research},'' {Accessed: \today}.
  [Online]. Available: \url{https://www.nvidia.com/en-us/data-center/dgx-1/}
\BIBentrySTDinterwordspacing

\bibitem{nccl-github}
\BIBentryALTinterwordspacing
------, ``{Optimized Primitives for Collective Multi-GPU Communication},''
  {Accessed: \today}. [Online]. Available: \url{https://github.com/NVIDIA/nccl}
\BIBentrySTDinterwordspacing

\bibitem{summit}
\BIBentryALTinterwordspacing
{Oak Ridge National Laboratory}, ``{SUMMIT},'' {Accessed: \today}. [Online].
  Available: \url{https://www.olcf.ornl.gov/summit/}
\BIBentrySTDinterwordspacing

\bibitem{potluri2013efficient}
S.~Potluri, K.~Hamidouche, A.~Venkatesh, D.~Bureddy, and D.~K. Panda,
  ``{Efficient Inter-node MPI Communication Using GPUDirect RDMA for InfiniBand
  Clusters with NVIDIA GPUs},'' in \emph{Parallel Processing (ICPP), 2013 42nd
  International Conference on}, Oct 2013, pp. 80--89.

\bibitem{schmidhuber2015deep}
J.~Schmidhuber, ``{Deep Learning in Neural Networks: An Overview},''
  \emph{Neural networks}, vol.~61, pp. 85--117, 2015.

\bibitem{SennrichHB16}
\BIBentryALTinterwordspacing
R.~Sennrich, B.~Haddow, and A.~Birch, ``{Edinburgh Neural Machine Translation
  Systems for {WMT} 16},'' \emph{CoRR}, vol. abs/1606.02891, 2016. [Online].
  Available: \url{http://arxiv.org/abs/1606.02891}
\BIBentrySTDinterwordspacing

\bibitem{Shi:2014:HiPC}
R.~Shi, S.~Potluri, K.~Hamidouche, J.~Perkins, M.~Li, D.~Rossetti, and D.~K.
  Panda, ``{Designing Efficient Small Message Transfer Mechanism for Inter-node
  MPI Communication on InfiniBand GPU Clusters},'' in \emph{2014 21st
  International Conference on High Performance Computing (HiPC)}, Dec 2014, pp.
  1--10.

\bibitem{scatter-allgather-2}
M.~Shroff and R.~A. V.~D. Geijn, ``{CollMark: MPI Collective Communication
  Benchmark},'' Dept. of Computer Sciences, University of Texas at Austin,
  Tech. Rep., 2000.

\bibitem{vgg-paper}
K.~Simonyan and A.~Zisserman, ``{Very Deep Convolutional Networks for
  Large-Scale Image Recognition},'' \emph{arXiv preprint arXiv:1409.1556},
  2014.

\bibitem{googlenet}
C.~Szegedy, W.~Liu, Y.~Jia, P.~Sermanet, S.~Reed, D.~Anguelov, D.~Erhan,
  V.~Vanhoucke, and A.~Rabinovich, ``{Going Deeper with Convolutions},'' in
  \emph{Proceedings of the IEEE Conference on Computer Vision and Pattern
  Recognition}, 2015, pp. 1--9.

\bibitem{Thakur:2005:OCC}
R.~Thakur, R.~Rabenseifner, and W.~Gropp, ``{Optimization of Collective
  Communication Operations in MPICH},'' \emph{Int. J. High Perform. Comput.
  Appl.}, vol.~19, no.~1, pp. 49--66, Feb. 2005.

\bibitem{openmpi}
{The Open MPI Development Team}, ``{Open MPI : Open Source High Performance
  Computing},'' http://www.open-mpi.org.

\bibitem{Venkatech:2014:HiPC:gpu-mcast}
A.~Venkatesh, H.~Subramoni, K.~Hamidouche, and D.~K. Panda, ``{A High
  Performance Broadcast Design with Hardware Multicast and GPUDirect RDMA for
  Streaming Applications on Infiniband Clusters},'' in \emph{2014 21st
  International Conference on High Performance Computing (HiPC)}, Dec 2014, pp.
  1--10.

\bibitem{Zhou:2015:ICPPW}
H.~Zhou, V.~Marjanovic, C.~Niethammer, and J.~Gracia, ``{A Bandwidth-Saving
  Optimization for MPI Broadcast Collective Operation},'' in \emph{2015 44th
  International Conference on Parallel Processing Workshops}, Sept 2015, pp.
  111--118.

\end{thebibliography}

%\begin{thebibliography}{1}


%\end{thebibliography}

% Generated by IEEEtranS.bst, version: 1.14 (2015/08/26)

\end{document}